\documentclass[aps,prb,twocolumn,superscriptaddress,showpacs]{revtex4-1}
\usepackage{graphicx}
\usepackage{dcolumn}
\usepackage{bm}
\usepackage{subfigure}
\usepackage{amsmath}
\usepackage{amssymb}
\usepackage{hyperref} 
\usepackage{xcolor}

\newcommand{\ua}{\uparrow} 
\newcommand{\da}{\downarrow} 
\newcommand{\ra}{\rightarrow}

\newcommand{\vk}{{\vec{k}}} 
\newcommand{\bs}{\boldsymbol}
\newcommand{\SRO}{Sr$_2$RuO$_4$}

\usepackage{times} 

\begin{document} 
\title{A review of some new perspectives on the theory of superconducting \SRO}
\author{Wen Huang}
\email{huangw3@sustech.edu.cn}
\address{Shenzhen Institute for Quantum Science and Engineering, Southern University of Science and Technology, Shenzhen 518055, Guangdong, China}

\date{\today}

\begin{abstract}
The nature of the Cooper pairing in the paradigmatic unconventional superconductor \SRO~is an outstanding puzzle in condensed matter physics. Despite the tremendous efforts made in the past twenty-seven years, neither the pairing symmetry nor the underlying pairing mechanism in this material has been understood with clear consensus. This is largely due to the lack of a superconducting order that is capable of interpreting in a coherent manner the numerous essential experimental observations. At this stage, it may be desirable to reexamine the existing theoretical descriptions of superconducting \SRO. This review focuses on several recent developments that may provide some clues for future study. We highlight three separate aspects: 1) any pairing in the $E_u$ symmetry channel, with which the widely discussed chiral p-wave is associated, shall acquire a 3D structure due to spin-orbit entanglement; 2) if the reported Kerr effect is a superconductivity-induced intrinsic bulk response, the superconductivity must either exhibit a chiral character, or be complex mixtures of certain set of helical p-wave pairings; 3) when expressed in a multiorbital basis, the Cooper pairing could acquire numerous exotic forms that are inaccessible in single-orbital descriptions. The implications of each of these new perspectives are briefly discussed in connection with selected experimental phenomena. 
\end{abstract}

\maketitle
Following the discovery of the high-temperature superconductivity in cuprates, an effort to look for superconducting layered perovskites that do not contain copper ended in \SRO~\cite{Maeno:94}. This material features a relatively low transition temperature $T_c\sim 1.5$ K in the best available samples, and the unconventional nature of the pairing is evident in its extreme sensitivity to disorder~\cite{Mackenzie:98}. From the start, it was noted that it most likely does not follow the same paradigm as the cuprates. The first proposal was spin-triplet odd-parity p-wave pairing analogous to that in the superfluid Helium-3~\cite{Rice:95,Baskaran:96}. Initial support came in an NMR Knight shift measurement~\cite{Ishida:98} and separately in a polarized neutron experiment~\cite{Duffy:00}, which suggested an unchanged spin susceptibility measured parallel to the RuO$_2$ plane across the superconducting transition. The odd-parity nature was later substantiated by a Josephson interference measurement~\cite{Nelson:04}. In view of the further reports of time-reversal symmetry breaking in $\mu$SR~\cite{Luke:98} and optical polar Kerr effect measurements~\cite{Xia:06}, this material became an almost perfect example of a chiral $p_x+ip_y$ superconductor, similar to the 2D version of the Helium-3 A-phase~\cite{Leggett:75}. This makes \SRO~a potential topological superconductor suitable for topological quantum computation~\cite{Read:00,Ivanov:01}, which has largely driven the research in the field before alternative material systems rose to prominence in the recent decade.

\begin{figure}
\includegraphics[width=7cm]{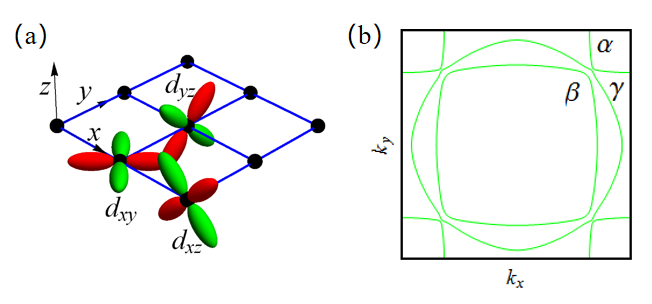}
\caption{(a) The band structure of \SRO~near the Fermi level is typically simulated by a three-orbital model, with Ru $d_{xz}$, $d_{yz}$ and $d_{xy}$ orbitals residing on each site of a square lattice. (b) Three Fermi surfaces obtained from a representative three-orbital model.}
\label{fig:lattice}
\end{figure}

The potential topological property put aside, \SRO~probably constitutes one of the best solid state material platforms where both the mechanism and the detailed form of the unconventional Cooper pairing can be exactly solved. Its fermiology~\cite{Mackenzie:03}, including the electronic structure and the relatively simple Fermi surface geometry (Fig.~\ref{fig:lattice})~\cite{Bergemann:00,Bergemann:03,Damascelli:00,Haverkort:08,Veenstra:14}, has been characterized with unprecedented accuracy --- an incredible feat not often possible in other unconventional superconductors. Further, while effects of electron correlations are noticeable in the normal state, superconductivity actually emerges out of a well-behaved Fermi liquid, featuring a standard Bardeen-Schrieffer-Cooper type transition~\cite{Mackenzie:03}. These led to a popular perception that the problem shall be well within the reach of established theoretical and experimental techniques.

However, despite the seeming early success, the jigsaw has never been completed. The controversy is mainly driven by numerous observations in the course of the past two decades that contradict the expected behavior of a standard chiral p-wave. For example, thermodynamic and transport studies point to a nodal quasiparticle excitation spectrum~\cite{Ishida:00,NishiZaki:00,Bonalde:00,Tanatar:01,Izawa:01,Lupien:01,Deguchi:04,Firmo:13,Hassinger:17}, as opposed to a full gap typically expected for chiral p-wave; scanning probes have failed to detect the expected spontaneous current at sample edges~\cite{Kirtley:07,Hicks:10,Curran:14}; there has been no thermodynamic evidence of split superconducting transitions under perturbations that break the degeneracy between the $p_x$ and $p_y$ order parameter components~\cite{Yonezawa:14,Hicks:14,Steppke:17,LiYS:21}; the in-plane upper critical field appears to be Pauli-limited~\cite{Yonezawa:13,Kittaka:14,Yonezawa:14}, a fact that aligns more fittingly with a spin-singlet rather than a spin-triplet pairing. While some of these discrepancies permit explanations within the chiral p-wave framework by invoking microscopic details or sample imperfection, the final straw came in 2019. A revised Knight shift measurement~\cite{Pustogow:19}, along with quick subsequent confirmations~\cite{Ishida:20,Petsch:20}, pointed to a strongly suppressed spin susceptibility below $T_c$, constituting by far the strongest evidence against the presumed chiral p-wave order. A number of new developments~\cite{Sharma:20,Grinenko:21,Benhabib:21,Ghosh:21,Chronister:21,Grinenko:21b} followed since then, each bringing significant new knowledge of \SRO. However, till this date, no one seems able to speak with confidence that any specific superconducting order coherently interprets all of the key observations. The current dilemma therefore seems to necessitate a critical reexamination of certain subset of experiments or their theoretical interpretations. 

Here, we do not seek to provide a thorough overview of how various order parameter symmetries may or may not interpret certain essential signatures. In fact, the said task has been nicely conducted at multiple stages during the last two decades. Interested readers are referred to Refs.~\onlinecite{Maeno:01,Mackenzie:03,Sigrist:05,Kallin:09,Maeno:12,Kallin:12,Liu:15,Kallin:16,Mackenzie:17,Wysokinski:19,Leggett:20,Anwar:21} for more details. Our primary objective is to highlight several recent developments, which we believe bring some new perspectives and may provide some clues for future studies. We cover the following three independent aspects:
\begin{itemize}
\item If \SRO~does stabilize a superconducting order in the $E_u$ symmetry channel, the pairing shall inevitably acquire a 3D structure due to spin-orbit coupling. This could lead to nontrivial consequence on the mean-field ground state, i.e.~chiral versus nematic p-wave order. (See Ref.~\onlinecite{Huang:18})
\item If the Kerr effect is a superconductivity-induced clean-limit bulk response, the superconducting state shall exhibit either a chiral pairing (e.g. $p+ip$, $d+id$, etc) or a complex mixture of certain helical p-wave pairings. In addition, the Kerr measurement may not resolve the distribution of the pairing among the multiple orbitals/bands. (See Refs.~\onlinecite{ZhangJL:20,Huang:21})
\item A multi-orbital description permits numerous exotic forms of Cooper pairing, which may lead to properties unavailable in single-orbital models. (See Refs.~\onlinecite{Huang:19,Ramires:19,Kaba:19})
\end{itemize}
In one way or another, these aspects have not received due attention in the past. 

Before we proceed, it is important to stress that these studies in no way encompass all of the recent theory advances the community has made towards unraveling the mystery in \SRO.

\section{A consensus: multi-component order parameter}
It is worth first pointing out what is perhaps the most widely accepted view, that \SRO~condenses at least two superconducting order parameter components. This has been inferred from a variety of experiments, including $\mu$SR~\cite{Luke:98,Grinenko:21}, optical polar Kerr effect~\cite{Xia:06}, Josephson interferometry~\cite{Kidwingira:06,Anwar:13,Saitoh:15}, and ultrasound~\cite{Ghosh:21,Benhabib:21}, etc. One possibility is a pairing in a multi-dimensional irreducible representation (irrep). A representative example is the well-known chiral p-wave pairing with gap function $(k_x+ik_y)\hat{z}$, which belongs to the $E_u$ irrep of the $D_{4h}$ group. Here, $\hat{z}$ denotes the orientation of the so-called d-vector, which describes the spin configuration of a spin-triplet pairing~\cite{Mackenzie:03}. Another possibility is coexisting order parameters associated with distinct one-dimensional irreps. In this scenario, the multiple components typically condense below distinct critical temperatures, although accidental near-degeneracy is possible, in principle. On the other hand, one cannot rule out the possibility that a single-component order parameter is condensed in clean samples, while other coexisting subdominant components emerge around impurities or crystalline dislocations. 

Take the example of a simple single-orbital model, a general multi-component gap function can be written as, 
\begin{equation}
\hat{\Delta}_{\vk} = \left[\sum_m \phi_m^s f_{m,\vk} + \sum_n \phi_n^t \vec{d}_{n,\vk}\cdot \vec{s} \right] is_y  \,.
\end{equation}
Here, $\phi_m^s$ and $\phi_n^t$ stand respectively for the order parameter components of spin-singlet and spin-triplet pairings, while $f_{m,\vk}$ and $\vec{d}_{n,\vk}$ depict the forms of the pairings in the corresponding symmetry channels. 

\section{A general $E_u$ state}
\label{sec:1}
Table \ref{tab:tab1} lists all of the five possible odd-parity pairing channels for \SRO. For a long time, the $E_u$ pairing with $(k_x\hat{z},k_y\hat{z})$ dominated the narratives. The spatial part of this Cooper pair wavefunction lies in the $xy$ plane. However, there is in fact an out-of-plane pairing, $(k_z\hat{x},k_z\hat{y})$, which shares the same symmetry and has received far less attention. Given the enormous challenge the former faces at present, it is worth exploring the new physics brought about by this often-ignored ingredient. In this section, we argue for the general presence of the $k_z$-like pairing in the $E_u$ symmetry channel and discuss its consequences~\cite{Huang:18}. 

\begin{table}[t]
\caption{\label{tab:table1} Superconducting gap functions in the five odd-parity irreducible representations of the $D_{4h}$ group. The pairings in the $A_{iu}$ and $B_{iu}$ symmetry channels are often referred to as helical p-wave pairings, while that in the $E_u$ channel are typically associated with chiral or nematic p-wave pairing.}
{\renewcommand{\arraystretch}{1.3}
\begin{tabular}{c|c}
\hline
~~~~irrep.~~~~ & ~~~~~~~~~basis function ($\vec{d}_{\bs k}$)~~~~~~~\\
\hline
$A_{1u}$                &     $k_x \hat{x} + k_y \hat{y}$; $k_z\hat{z}$  \\
\hline
$A_{2u}$                    &     $k_y \hat{x} - k_x \hat{y}$  \\ \hline
$B_{1u}$     &     $k_x\hat{x}-k_y\hat{y}$                \\ \hline
$B_{2u}$    &     $k_y\hat{x} + k_x \hat{y}$                                  \\ \hline
$E_{u}$       &     $(k_x,k_y)\hat{z}; ~~(k_z \hat{x}, k_z \hat{y})$ \\ \hline
\end{tabular}}
\label{tab:tab1}
\end{table}

We begin by noting that, in the presence of SOC, spins are no longer good quantum numbers. Hence the spin and spatial parts of the Cooper pair wavefunctions are entangled, and symmetry operations must act simultaneously on both parts. One can check that the two types of pairings mentioned above behave the same way under any $D_{4h}$ operation. For example, a $C_4$ rotation turns $(k_x\hat{z},k_y\hat{z})$ to $(k_y\hat{z},-k_x\hat{z})$, and similarly $(k_z\hat{x},k_z\hat{y})$ to $(k_z\hat{y},-k_z\hat{x})$; while a mirror reflection about the $yz$-plane transforms $(k_x\hat{z},k_y\hat{z})$ to $(k_x\hat{z},-k_y\hat{z})$, and $(k_z\hat{x},k_z\hat{y})$ to $(k_z\hat{x},-k_z\hat{y})$; etc. Note that the spin part of the wavefunction (i.e. $\hat{x}$, $\hat{y}$, $\hat{z}$) transforms like a pseudovector. 

The in-plane pairing forms the commonly discussed chiral p-wave state when the two components $k_x\hat{z}$ and $k_y\hat{z}$ develop a phase difference of $\pi/2$ or $-\pi/2$. On the other hand, if the two develop a phase difference of 0 or $\pi$ or, if only one component is condensed, the resultant state is a nematic p-wave that breaks in-plane rotation symmetry. The out-of-plane basis $(k_z \hat{x}, k_z \hat{y})$ has often been ignored in the discussion of \SRO, thanks to its quasi-2D electronic structure around the Fermi level. However, there are indeed finite interlayer couplings, which generate a 3D variation of spin-orbit entanglement~\cite{Veenstra:14}. This suggests that the two symmetry-equivalent bases shall generically coexist, and a general form of the $E_u$ pairing would be,
\begin{equation}
(\vec{d}_{1,\bs k},\vec{d}_{2,\bs k}) = (k_x \hat z + \epsilon k_z \hat x, k_y \hat z+\epsilon k_z \hat y)\,,
\label{eq:newD}
\end{equation}
where $\epsilon$ is a non-universal real constant determined by the relative strength of the effective interactions responsible for the respective in-plane and out-of-plane pairings. In other words, the $E_u$ pairing in \SRO~shall be 3D. 

Including the out-of-plane pairing has a nontrivial consequence on the selection of the superconducting ground state, which we now analyze through a Ginzburg-Landau (GL) theory. One may write the generic two-component $E_u$ gap function as $\hat{\Delta}_{\bs k} =( \phi_1 \vec{d}_{1,\bs k}\bs\cdot \vec{s} + \phi_2 \vec{d}_{2,\bs k}\bs\cdot \vec{s})is_y$, where $\phi_{1(2)}$ label the order parameter components associated with $\vec{d}_{1(2),\bs k}$. The GL free energy, up to the quartic order, reads,
\begin{eqnarray}
f&=& r(T-T_c)\left(|\phi_1|^2 + |\phi_2|^2\right) + \beta \left(|\phi_1|^4+|\phi_2|^4 \right) \nonumber \\
&&+ \beta_{12}|\phi_1|^2|\phi_2|^2 + \beta^\prime \left(\phi_1^\ast\phi_2 + \phi_1\phi_2^\ast \right)^2 \,.
\label{eq:GLaction}
\end{eqnarray}
Within mean-field theory, it can be shown that the stability of the phases is determined by the sign of $\beta^\prime$. If $\beta^\prime>0$, $\phi_1$ and $\phi_2$ preferentially develop a $\pi/2$ phase difference, i.e. $\bs{\phi}= (\phi_1,\phi_2)=\phi(1,\pm i)$, thereby breaking time-reversal invariance (TRI), as for the chiral p-wave order; whilst if $\beta^\prime<0$, the system favors a time-reversal invariant order parameter, $\bs{\phi} = \phi(1, \pm 1)$ or $\phi(1,0)$. The latter breaks rotational symmetry and is often referred to as nematic pairing.  

\begin{figure}[t]
\includegraphics[width=5.cm]{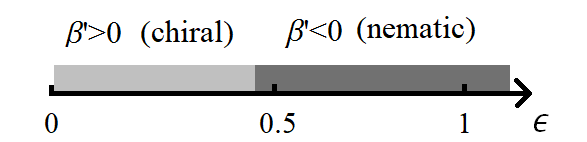}
\caption{(Reproduced with permission from Ref.~\onlinecite{Huang:18}). The phase diagram as a function of $\epsilon$ for an $E_u$ pairing given by (\ref{eq:newD}). The chiral and nematic phases are shaded in light and dark grey, respectively. We assume cylindrical Fermi surface with radius 1 and replace $k_z$ by $\sin k_z$ (taking the range of $k_z$ to be $[-\pi,\pi]$ in the integral). In this calculation $\beta^\prime$ changes sign at $\epsilon_c\approx 0.46$. }
\label{fig:phaseDiag}
\end{figure}

It turns out $\beta^\prime$ relates to the structure of $\vec{d}_{i,\bs k}$ through the following approximation,
\begin{equation}
\beta^\prime \approx C \left\langle  (\vec{d}_{1,\bs k}\cdot\vec{d}_{2,\bs k})^2 - |\vec{d}_{1,\bs k} \times \vec{d}_{2,\bs k} |^2  \right \rangle_\text{FS} \,,
\label{eq:betaPtext}
\end{equation}
where $C$ is a positive constant and $\langle \cdots \rangle_\text{FS}$ denotes an integral over the Fermi surface. Plotted in Fig.~\ref{fig:phaseDiag} is the phase diagram as determined by the sign of $\beta^\prime$ in a simple case of a cylindrical Fermi surface. We see that $\beta^\prime$ in turn depends on the coefficient $\epsilon$. For the considered model, an out-of-plane pairing slightly less than half the amplitude of the in-plane pairing is sufficient to turn the favored state from chiral to nematic. 

Even within a simple single-band description, a 3D $E_u$ pairing has certain advantage over its 2D counterpart. In particular, the presence of $k_z$-like pairing which has in-plane d-vector orientations makes possible a generic suppression in the in-plane spin susceptibility below $T_c$. For the nematic pairing, the degree of suppression differs for different components of the susceptibility. This readily alludes to the drop of spin susceptibility below $T_c$~\cite{Pustogow:19,Ishida:20,Petsch:20}, although any serious connection must be made in conjunction with a more realistic multiorbital spin-orbit coupled model for \SRO. 

Between the chiral and nematic $E_u$ pairings, a nematic state stands a better chance to explain several other outstanding puzzles. Foremost, being time-reversal invariant, it easily explains the absence of spontaneous surface current~\cite{Kirtley:07,Hicks:10,Curran:14}. Secondly, while a chiral $E_u$ state expects a sequence of two superconducting transitions under in-plane uniaxial stress or magnetic field, a nematic state may see only a single transition as reported~\cite{Yonezawa:14,Hicks:14,Steppke:17,LiYS:21}. Such perturbations break the degeneracy between the two $E_u$ components. For the chiral state, the two transitions break distinct symmetries: the $U(1)$ symmetry at the upper transition, and time-reversal symmetry at the lower one. For the nematic state with order parameter $\phi(1,0)$, no additional transition occurs if the stated perturbations are applied along the crystalline $x$- or $y$-axis. The other nematic state with $\phi(1,1)$ will exhibit a lower transition at which a reflection symmetry about the vertical plane parallel to the $x$- or $y$-direction is broken, but only when the stress or external field aligns exactly with these crystalline axes. Any misalignment inadvertently breaks the reflection symmetry, causing the second transition to smear out. A nematic state is not unchallenged though, given the unambiguous signatures of time-reversal symmetry breaking below $T_c$~\cite{Luke:98,Xia:06}. A possible explanation is local chiral pairings nucleated around impurities or lattice dislocations, while the bulk favors a nematic pairing in the clean limit.  

\section{Implications of intrinsic Kerr effect}
A prominent observation in superconducting \SRO~is the optical polar Kerr effect, in which a linearly polarized light normally incident upon the sample is reflected with a rotated polarization. This peculiar effect implies finite a.c. anomalous Hall conductivity, and it reveals crucial information about the symmetry of the superconducting order parameter. Attempts to explain this phenomenon invoked either extrinsic or intrinsic mechanisms~\cite{Goryo:08,Lutchyn:09,Taylor:12,Wysokinski:12,Konig:17,Komendova:17,LiY:20,ZhangJL:20,Denys:21}. In the following, we focus on the implication of Kerr effect under the assumption that it is a genuine response of the superconducting state in the clean limit, i.e.~not related to defects or other forms of sample imperfection. That is, we ask what kind of bulk superconducting order could support intrinsic Kerr effect. 

\subsection{Pairing symmetry}
As the Kerr rotation is closely related to the Hall effect, it is natural, by analogy with quantum (anomalous) Hall effects, to expect the underlying Cooper pairs to exhibit finite orbital angular momentum, such as in the form of chiral $p+ip$ and $d+id$ pairings. This leads us to two distinct scenarios. One is a genuine chiral superconducting order, and the other is overall non-chiral but effectively consists of copies of chiral pairings with opposite chirality and inequivalent gap functions. 

The most frequently discussed pairings in the first scenario includes the $p_x+ip_y$ (including its 3D generalization discussed in Sec.~\ref{sec:1}), $d_{xz}+id_{yz}$ and $d_{x^2-y^2}+id_{xy}$ states, which belong to the $E_u$, $E_g$, and mixed $B_{1g}$ and $B_{2g}$ irreps, respectively. Similar to the $p_x+ip_y$ state, the other two states both suffer some shortfalls in regards to their compatibility with experiments. The $d_{xz}+id_{yz}$ state, although consistent with the discontinuity in the shear elastic modulus $c_{66}$ in ultrasound spectroscopy~\cite{Benhabib:21,Ghosh:21}, generally produces finite spontaneous current~\cite{Nie:20}. Further, two consecutive superconducting transitions are expected in the presence of in-plane perturbations, but no thermodynamic evidence has been reported~\cite{Yonezawa:14,Hicks:14,Steppke:17,LiYS:21}. The $d_{x^2-y^2}+id_{xy}$ state generates vanishingly small current~\cite{Huang:14,Tada:15}, although it would require an accidental degeneracy for the $B_{1g}$ and $B_{2g}$ components to converge to a single superconducting transition in an unperturbed crystal, and in the clean limit it cannot trigger a discontinuity in $c_{66}$~\cite{Huang:21}.   

\begin{figure}[htp]
\includegraphics[width=8.5cm]{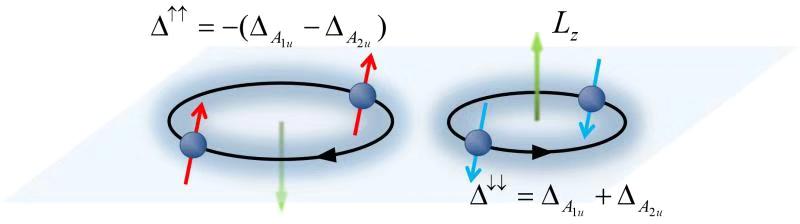}
\caption{(Adapted from Ref.~\onlinecite{Huang:21}). Illustration of the Cooper pairing in the mixed helical p-wave $A_{1u}+iA_{2u}$ state in a single-orbital model. The state effectively consists of two subsets of (spinless) chiral p-wave models exhibiting opposite chirality. The Cooper pairs of the two subsets are drawn with different size to reflect their different gap amplitudes. Green arrows indicate the direction of the Cooper pair orbital angular momentum $L_z$. In the limit $\Delta_{A_{1u}}= \Delta_{A_{2u}}$, the state resembles the $^3$He A$_1$-phase. Another candidate state $B_{1u}+iB_{2u}$ has a similar pairing configuration. }
\label{fig:mixedHelical}
\end{figure}

The second scenario is rarely recognized. Specific to candidate two-component pairings, there are two possibilities: non-unitary $A_{1u}+iA_{2u}$ and $B_{1u}+iB_{2u}$ pairings~\cite{Huang:21}. It can be checked that no other two-component non-chiral but time-reversal symmetry breaking state conceivable for \SRO~supports intrinsic Kerr effect. The $A_{iu}$ and $B_{iu}$ pairings are traditionally referred to as helical p-wave pairings (see Table \ref{tab:tab1}). Each of the resultant mixed helical states can in fact be viewed as a product of two subsets of `spinless' chiral superconductors with opposite chirality and different pairing amplitudes (e.g.~Fig.~\ref{fig:mixedHelical}). Hence, the Hall conductivity receives opposite yet not fully canceled contributions from the two subsets. Similar, and more substantial cancellation, could be found for the spontaneous surface current. The in-plane d-vector also paves way for a drop of spin susceptibility below $T_c$, although a direct comparison to experiment requires more careful examination. 

Note that the setup of the Kerr measurement of \SRO~\cite{Xia:06} necessarily implies a superconducting state without any reflection symmetry about the mirror planes perpendicular to the $xy$-plane. However, the breaking of these mirror symmetries, combined with broken time-reversal symmetry, do not constitute a sufficient condition for finite Kerr rotation. One example is the mixed-parity $A_{1g}+iA_{1u}$ state --- a mixture of an $s$-wave and a helical $p$-wave order parameter. It has no vertical mirror symmetry, yet the Hall conductivity vanishes, irrespective of the underlying microscopic model details. 

\subsection{Dominant superconducting orbital(s)}
Since early days~\cite{Agterberg:97,Zhitomirsky:01} it was pointed out that superconductivity is most likely dominated by one set of the three orbitals, given the disparity between dispersion of the quasi-1D and quasi-2D orbitals. By interorbital proximity, pairing can also develop on the subdominant orbital(s) with a weaker strength. However, no consensus is available as to which set of orbital(s) dominates the pairing. The van Hove singularity on the $\gamma$-band ($d_{xy}$ orbital) and the enhanced spin fluctuations associated with the quasi-1D bands ($d_{xz}$ and $d_{yz}$ orbitals) due to quasi-nesting were separately argued to promote pairing instabilities of some kind~\cite{Takimoto:00,Nomura:00,Nomura:02,Eremin:02,Raghu:10,Huo:13,WangQH:13,Scaffidi:14,Tsuchiizu:15,ZhangLD:18,WangWS:19,Gingras:19,Romer:19}.

Previously, intrinsic Kerr effect was also regarded as a means to indirectly probe the microscopic distribution of the Cooper pairing on the multiple electron orbitals or bands in this material. On general grounds, clean single-orbital chiral superconductors do not support intrinsic Hall effect~\cite{Read:00}. This can be understood from the Galilean invariance principle: the anomalous Hall effect is associated with the center-of-mass motion of the Cooper pairs under an external electric field (in the absence of external magnetic field), which is unrelated to the relative motion between paired electrons. Hence the Cooper pair chirality, i.e.~the orbital angular momentum, does not automatically imply Hall effect. Under this pretext, the notion of pure $d_{xy}$-orbital-driven superconductivity in \SRO~is questioned. 

A multiorbital/multiband model, on the other hand, is non-Galilean-invariant, as the Cooper pair center-of-mass motion and relative motion now become entangled. The connection between the Kerr effect and the multiorbital superconductivity in \SRO~was first recognized in 2012~\cite{Taylor:12,Wysokinski:12}, based on a model with chiral pairing on the two quasi-1D orbitals. This development lends support to the alternative view that superconductivity is predominantly driven by the quasi-1D orbitals~\cite{Raghu:10}. However, a recent generalization of the above study to models with single-$d_{xy}$-dominated chiral pairing was also shown to support intrinsic Kerr effect~\cite{ZhangJL:20}. Those models took into account proximitized interorbital pairing between $d_{xy}$ and other orbitals, e.g.~the quasi-1D orbitals or, the oxygen $p$-orbitals that have been customarily ignored in most studies. The resultant Kerr rotation, calculated based on reasonably realistic estimate of the pairing gaps, equally well accounts for the experimental observation. In brief, intrinsic Kerr effect may not resolve the distribution of superconducting gap on the multiple d-orbitals.

\begin{table*}
\caption{(Reproduced with permission from Ref.~\onlinecite{Huang:19} with minor revision). Representative basis functions of the superconducting pairing in the two-orbital model in various irreps of the $D_{4h}$ point group. Here $\sigma_i$ and $s_i$ $(i=x,y,z)$ are Pauli matrices that operate in the orbital and spin space, respectively, while $\sigma_0$ and $s_0$ are the corresponding identity operators. The vectors $\hat{x}$, $\hat{y}$ and $\hat{z}$ denote the components of the d-vector of spin-triplet pairings, and the pairing gap functions are obtained by multiplying the basis function by $is_y$. Note that we have neglected out-of-plane pairings in the table.}
{\renewcommand{\arraystretch}{1.5}
\begin{tabular}{l c}
\hline
\hline
irrep.~~~~ & ~~~~~~~~~basis function ~~~~~~~\\
\hline
$A_{1g}$   & $\hat{I}$,~~$i\sigma_y\otimes \hat{z}\!\cdot\! \vec{s} $,~~$k_xk_y \sigma_x $,~~$(k_x^2-k_y^2)\sigma_z$ \\
\hline

$A_{2g}$   &  $k_xk_y\sigma_z $,~~$(k_x^2-k_y^2)\sigma_x $  \\
\hline

$B_{1g}$   &  $\sigma_z $,~~$i\sigma_y\otimes (k_x^2-k_y^2)\hat{z} \!\cdot\! \vec{s} $  \\
\hline

$B_{2g}$   &  $\sigma_x $,~~ $i\sigma_y\otimes k_xk_y\hat{z} \!\cdot\! \vec{s} $  \\
\hline

$E_g$       &   $\left( i\sigma_y\otimes  \hat{x} \!\cdot\! \vec{s} ,~~i\sigma_y\otimes \hat{y} \!\cdot\! \vec{s}  \right)$ \\

\hline

$A_{1u}$   & {$\frac{\sigma_0\pm \sigma_z}{2}\otimes k_x\hat{x} \!\cdot\! \vec{s}+ \frac{\sigma_0\mp \sigma_z}{2}\otimes k_y\hat{y} \!\cdot\! \vec{s} $,~~ $\sigma_x\otimes (k_x \hat{y} + k_y \hat{x})\!\cdot\! \vec{s} $}  \\
\hline

$A_{2u}$   & $\frac{\sigma_0\pm \sigma_z}{2}\otimes k_x\hat{y} \!\cdot\! \vec{s}- \frac{\sigma_0\mp \sigma_z}{2}\otimes k_y\hat{x} \!\cdot\! \vec{s} $,~~ $\sigma_x\otimes (k_x \hat{x} - k_y \hat{y})\!\cdot\! \vec{s} $  \\
\hline

$B_{1u}$   & $\frac{\sigma_0\pm \sigma_z}{2}\otimes k_x\hat{x} \!\cdot\! \vec{s} - \frac{\sigma_0\mp \sigma_z}{2}\otimes k_y\hat{y} \!\cdot\! \vec{s}$,~~ $\sigma_x\otimes (k_x \hat{y} - k_y \hat{x})\!\cdot\! \vec{s} $  \\
\hline

$B_{2u}$   & $\frac{\sigma_0\pm \sigma_z}{2}\otimes k_x\hat{y} \!\cdot\! \vec{s}+ \frac{\sigma_0\mp \sigma_z}{2}\otimes k_y\hat{x} \!\cdot\! \vec{s}$,~~ $\sigma_x\otimes (k_x \hat{x} + k_y \hat{y})\!\cdot\! \vec{s}$  \\
\hline
\shortstack{ ~~ \\ $E_u$ \\ ~~ \\ ~~\\ ~~\\ ~~}        &

\shortstack{~~\\ $\left( i k_x \sigma_y,~~i k_y\sigma_y  \right)$\\
~~\\
$\left( \sigma_x\otimes k_y \hat{z}\!\cdot\!\hat{s} ,~~\sigma_x \otimes k_x\hat{z}\!\cdot\!\vec{s} \right)$  \\
~~\\
$ \left(\frac{\sigma_0 \pm \sigma_z}{2}\otimes k_x\hat{z}\!\cdot\!\vec{s},~~\frac{\sigma_0 \mp \sigma_z}{2}\otimes  k_y\hat{z}\!\cdot\!\vec{s} \right)$ }
\\
\hline
\end{tabular}}
\label{tab:tab2}
\end{table*}

\section{Symmetry classification in a multi-orbital description}
\SRO~has three Fermi sheets derived mainly from the Ru $4d$ $t_{2g}$-orbitals \cite{Damascelli:00,Bergemann:00} (Fig.~\ref{fig:lattice}). As superconductivity appears to emerge out of a coherent Fermi liquid and the critical temperature is relatively low, weak-coupling theory is widely considered applicable~\cite{Mackenzie:03}. In this setting, electrons near the Fermi level are considered most relevant to Cooper pairing, and it is usually assumed that only intraband pairing takes place. The (intraband) gap classification is relatively straightforward and has been well documented~\cite{Sigrist:91}. In the presence of finite SOC, spins are no longer good quantum numbers. Nonetheless, thanks to the Kramers degeneracy on the bands, an effective pseudospin basis can be adopted~\cite{Scaffidi:14}, and the same classification scheme follows. 

An alternative perspective is the orbital-basis description, wherein the Cooper pairs are formed between electrons with well-defined orbital characters. In this case, the usual classifications into even-parity spin-singlet and odd-parity spin-triplet pairings are no longer complete. One must also consider Cooper pairs symmetric and anti-symmetric in the orbital manifold~\cite{DaiX:08,ZhouY:08,WanY:09}. On top of this is further complexities introduced by interorbital hybridization and SOC, as we shall show below. The consequence is a rich variety of Cooper pairings that are inaccessible in single-orbital systems. While some aspects of this physics were already discussed in, e.g. Refs.~\onlinecite{Puetter:12,Gingras:19,WangWS:19}, more systematic classification and analyses were carried out in Refs.~\onlinecite{Huang:19,Ramires:19,Kaba:19}.  

In the following, we illustrate the essential physics using a pedagogical two-orbital model with $d_{xz}$ and $d_{yz}$ orbitals. A generic $\vk\cdot\vec{p}$ Hamiltonian that respects both time-reversal and the underlying $D_{4h}$ point group symmetries reads, in the spinor basis $(c_{xz\ua},c_{xz\da},c_{yz\ua},c_{yz\da})^T$,
\begin{equation}
H_{0\bs k}= t(k_x^2+k_y^2)-\mu+ \tilde{t}(k_x^2-k_y^2)\sigma_z + t^{\prime\prime}k_xk_y\sigma_x + \eta\sigma_y \otimes s_z  \,,
\label{eq:H0}
\end{equation}
Here $\sigma_i$ and $s_i$ with $i=x,y,z$ are the Pauli matrices operating on the respective orbital and spin degrees of freedom, $(t,\tilde{t},t^{\prime\prime})$ designate the kinetic energy and $\eta$ the onsite spin-orbit coupling (SOC). 

It is instructive to first analyze how the individual terms in Eq. \ref{eq:H0} respects $D_{4h}$. Most crucially, it must be recognized that the point group operations must act jointly on spatial, spin and orbital degrees of freedom. For example, a $C_4$ rotation, in addition to rotating momentum and spin, also exchanges the label of the two orbitals and induces a $\pi$ phase change on one of them, e.g. $(d_{xz},d_{yz}) \ra (d_{yz},-d_{xz})$. As a consequence, the bilinear $\sigma$-operators, which are formally $c^\dagger_{m,s}\sigma_i^{mn} c_{n,s^\prime}$ $(m=xz,yz)$, transform according to irreps of $D_{4h}$ in the following fashion: $\sigma_0$, $\sigma_x$, $\sigma_z$ and $\sigma_y$ as $A_{1g}$, $B_{2g}$, $B_{1g}$ and $A_{2g}$, respectively. A Hamiltonian invariant under all $D_{4h}$ operations is then constructed by appropriate product of the $\sigma$, $s$, and the momentum-space basis functions, as in Eq (\ref{eq:H0}). Note that, amongst the terms in the Hamiltonian, $s_z$ transforms as $A_{2g}$. Further, since the orbital wavefunction of the $t_{2g}$-electrons are even under inversion, the only effect of inversion is to invert electron momentum. This differs from the model with $p_x$ and $p_y$-orbitals, where inversion also changes the sign of the fermion creation and annihilation operators (the bilinear operators are however unaffected by this). 

The symmetry classification of the pairing gaps can be performed in a similar manner, with transformation properties of $\sigma$ and $s$ matrices essentially following the same pattern. Table \ref{tab:tab2} lists the representative superconducting basis functions in different irreps of the $D_{4h}$ group for our two-orbital model, where, for simplicity, we have only included in-plane pairings (i.e. omitting $k_z$-dependent terms). We see that most of the individual irreps contain multiple symmetry-equivalent basis functions, and some even have both spin-singlet and spin-triplet bases. These are prominent features not available in single-orbital systems. The individual pairings within each irrep cannot be distinguished by symmetry, and hence do not represent independent order parameters. In fact, they are generally brought into coexistence by orbital hybridization and/or SOC. In other words, in the case of generic orbital mixing, the multiple pairings belonging to the same symmetry irrep mutually induce one another. 

We take as an example a simple $A_{1g}$ gap function consisting of the first two bases in Table \ref{tab:tab2},
\begin{equation}
\hat{\Delta}_{\vk} = \psi_1 \hat{\Delta}_{1\vk} + \psi_2 \hat{\Delta}_{2\vk}  = ( \psi_1 \!\cdot\! \hat{I} + \psi_2 \!\cdot\!  i\sigma_y \otimes \vec{z}\!\cdot\! \vec{s} ) is_y \,.
\label{eq:A1gGap}
\end{equation}
Here, `$\hat{I}$' stands for $\sigma_0\otimes s_0$. The first and second terms correspond, respectively, to intra-orbital spin-singlet and inter-orbital spin-triplet pairings. The latter is also an orbital-triplet. The two do not necessarily coexist in the absence of SOC -- when spins are good quantum numbers. To see how SOC induces a mixing, we again resort to the Ginzburg-Landau theory. The singlet and triplet pairings are found to couple at quadratic order,
\begin{equation}
f_{12}= -i\lambda_{12} (\psi_1^\ast\psi_2 - \psi_2^\ast \psi_1)  \,,
\label{eq:f12SOC}
\end{equation}
with $\lambda_{12} \propto \eta$ a real constant. The complex phase is a consequence of the particular structure of the SOC in Eq.~\ref{eq:H0}. A similar conclusion was reached in Ref.~\onlinecite{Puetter:12}, and has also been argued from an experimental standpoint~\cite{Veenstra:14}. According to Eq.~(\ref{eq:f12SOC}), SOC not only mixes but also selects a particular relative phase between the two forms of pairing, e.g. $\theta_2-\theta_1= - \pi/2$ if $\lambda_{12}>0$. This relative phase can be absorbed into the basis function. Thus a more compact form of the gap function Eq.~\ref{eq:A1gGap} reads: $\hat{\Delta}_{\vec{k}} \propto \left[ \hat{I} + \mathcal{E} \sigma_y \otimes (\hat{z}\!\cdot\! \vec{s})  \right] i s_y$, where $\mathcal{E}$ is a real constant describing the relative amplitude between the two terms. Finally, it is easy to check that all of the four $A_{1g}$ pairings in Table~\ref{tab:tab2} are coupled, and therefore coexist, due to orbital mixing.

The natural spin-triplet and spin-singlet mixing calls for caution in regards to the interpretation of the spin susceptibility drop below $T_c$. More interestingly, had an $E_u$ pairing acquired a dominant form of $\left( i k_x \sigma_y,~~i k_y\sigma_y  \right)$ in Table~\ref{tab:tab2}, the pairing would be predominantly spin-singlet, in contrast to the traditional spin-triplet one. Another interesting observation is the existence of spin-triplet $E_g$ pairings in a pure 2D model (see Table~\ref{tab:tab2}), whereas in the conventional understanding it is a spin-singlet and 3D $(k_xk_z,k_yk_z)$-like pairing. 

Such exotic possibilities readily prompt new ideas for \SRO. Indeed, developments set to exploit the multiorbital nature have already been undertaken, with a multitude of attempts to reconcile with various experiments~\cite{LiYu:19,ChenWP:20,Suh:20,Lindquist:20,Clepkens:21}.

\section{Final remarks}
Having focused only on several works the author was involved in, this review is certainly limited in scope. Nevertheless, sound arguments were provided in each instance to justify the necessity to refine certain crucial aspects of our theoretical description of \SRO. Meanwhile, since our main purpose is to introduce these theory apparatuses, we have stopped short of combining these somewhat independent aspects to formulate a coherent interpretation of the ultimate puzzle in this material --- which lies beyond the capability of this brief review. 

Finally, this review is clearly not in a position to comment on the many other interesting and insightful recent theoretical proposals put forth by various groups, such as Refs.~\onlinecite{Ramires:16,WangZQ:20,Kivelson:20,Scaffidi:20,Willa:21,Romer:21,ZhangSJ:21,Clepkens:21} and many others unwittingly omitted here. These rich variety of proposals remind us of the very fact that we are nowhere near a full comprehension of the presumed unconventional Cooper pairing as a fundamental condensed matter phenomenon in what is perhaps one of the best studied superconducting materials. This is what, in part, molds the beauty of \SRO, as much as it brings about constant frustration to theorists.

\section{Acknowledgements}
The author would like to acknowledge extensive collaborations with Yu Li, Zhiqiang Wang, Hong Yao, Fu-Chun Zhang, Jia-Long Zhang and Yi Zhou, which led to the results that constitute the main body of this review. We are also grateful to Weipeng Chen, Wei-Qiang Chen, Ying Liu, Yongkang Luo, Catherine Kallin, Wenxing Nie, Aline Ramires, Thomas Scaffidi, Manfred Sigrist, Qiang-Hua Wang, Dao-Xin Yao, Fan Yang and Li-Da Zhang for many helpful discussions when those studies were carried out. This work is supported by NSFC under grant No.~11904155, the Guangdong Provincial Key Laboratory under Grant No.~2019B121203002, and a Shenzhen Science and Technology Program (KQTD20200820113010023).

\bibliography{SROreview_Sept23}
\end{document}